\documentclass[11pt,a4paper]{amsart}  %% FOR ARXIV E-PRINT 

\numberwithin{equation}{section}
\allowdisplaybreaks

\usepackage{amssymb}
\usepackage[latin1]{inputenc}
\usepackage{graphicx}
\usepackage{amsthm}
\usepackage{amsmath}
\usepackage{color}

\begin{document}

%\begin{frontmatter}




\title[C.M., Prabhakar function and non-Debye relaxation]{On complete monotonicity of
\\ the Prabhakar function 
\\ and non-Debye relaxation in dielectrics}
% \tnoteref{label1}}
%\tnotetext[label1]{The work of F.Mainardi has been carried out in the framework  of  %the GNFM - INdAM  activities. The work of R.Garrappa has been supported under %the GNCS - INdAM Project 2014.}

\author[F.Mainardi, R. Garrappa]{Francesco Mainardi$^{1}$, 
Roberto Garrappa$^{2}$}
%\corref{cor1}}
%\ead{francesco.mainardi@bo.infn.it}

\address{$^{1}$ Department of Physics and Astronomy, University of Bologna, and INFN \\ Via Irnerio 46, I-40126 Bologna, Italy}
\email{francesco.mainardi@bo.infn.it}

\address{$^{2}$ Department of Mathematics, University of Bari \\ Via E. Orabona 4, I-70125 Bari, Italy}
\email{roberto.garrappa@uniba.it}

\date{September 2016 
\\ %% MSC codes here, in the form: \MSC code \sep code
\\ 
{\it Note}: This E-Print is a revised version with a different layout
 of the paper published  in\\
{\bf  Journal of Computational Physics,
Vol. 293 (2015), pp. 70--80.} \\ DOI: 10.1016/j.jcp.2014.08.006
}

\begin{abstract}
The three parameters Mittag--Leffler function (often referred as the Prabhakar function) has important applications, mainly in physics of dielectrics, in describing anomalous relaxation of non--Debye type. This paper concerns with the investigation of the conditions, on the characteristic parameters, under which the function is locally integrable and completely monotonic; these properties are essential for the physical feasibility of the corresponding models. In particular the classical Havriliak--Negami model is extended to a wider range of the parameters. The problem of the numerical evaluation of the three parameters Mittag--Leffler function is also addressed and three different approaches are discussed and compared.  Numerical simulations are hence used to validate the theoretical findings and present some graphs of the function under investigation.
\end{abstract}

%\begin{keyword}

%\end{keyword}

%\end{frontmatter}
\maketitle 
%\linenumbers



\newtheorem{thm}{Theorem}
\newtheorem{lemma}[thm]{Lemma}
\newtheorem{proposition}[thm]{Proposition}
\newtheorem{rmk}{Remark}
%\newproof{proof}{Proof}

\def\alphaint{{\left\lceil \alpha \right\rceil}}
\def\Rset{{\mathbb{R}}}
\def\Cset{{\mathbb{C}}}
\def\Zset{{\mathbb{Z}}}
\def\Nset{{\mathbb{N}}}
\def\Ni{{N_{i}}}
\def\No{{N_{o}}}
\newcommand{\sgn}{\mathop{\rm sgn}}
\def\Arg{\mathop{\rm Arg}}
\def\e{\hbox {e}}
\def\dd{\hbox {d}}
\def\ds{\displaystyle}
\def\RR{\vbox {\hbox to 8.9pt {I\hskip-2.1pt R\hfil}}}
\def\NN{{\rm I\hskip-2pt N}}
\def\CC{{\rm C\hskip-4.8pt \vrule height 6pt width 12000sp\hskip 5pt}}

\section{Introduction}
\noindent
Recently Capelas de Oliveira, Mainardi and Vaz \cite{OliveiraMainardiVaz2011} 
have investigated the complete monotonicity of  the function of the Mittag-Leffler type %% of a real variable $t>0$ 
\begin{equation}\label{eq:ML3_Generalized}
	e_{\alpha,\beta}^{\gamma} (t) = t^{\beta-1} E_{\alpha,\beta}^{\gamma} ( - t^{\alpha}), \quad t\ge 0,
\end{equation}
where
\begin{equation}\label{eq:ML3_SeriesExpansion}
	 E_{\alpha,\beta}^{\gamma} (z) = \sum_{k=0}^{\infty} \frac{(\gamma)_{k} z^{k}}{k!\Gamma(\alpha k + \beta)}, \quad
	 (\gamma)_k= \frac{\Gamma(\gamma+k)}{\Gamma(\gamma)},
	 \quad z \in \CC, 
\end{equation}
denotes the Prabhakar function \cite{Prabhakar1971}
with  three positive order-parameters $\{\alpha, \beta, \gamma\}$.
For more details on this function see e.g. 
\cite{KilbasSrivastavaTrujillo2006,HauboldMathaiSaxena2011,%
Gorenflo-Kilbas-Mainardi-Rogosin2014}. 
It is also interesting to note that the Prabhakar function (\ref{eq:ML3_SeriesExpansion}) is a special case of the more general Fox--Wright function ${}_{1}\Psi_{1}$ which was first investigated by Wright in 1935 \cite{Wright1935}.

%{\color{red}{Recently, a fractional operator based on the Prabhakar function were investigated \cite{GarraGorenfloPolitoTomovski2014} with respect to applications in mathematical physics and probability, such us for instance for the description of the unsaturated behavior of the free electron laser or the derivation of a renewal point process generalizing the classical homogeneous and the time--fractional Poisson processes.}}

For some particular values of the parameters this function and its Laplace transform (LT)
\begin{equation}\label{eq:ML3_LaplaceTransform}
	{\mathcal E}_{\alpha,\beta}^\gamma (s) 
:= \int_0^\infty \!\e^{-st}\, e_{\alpha,\beta}^{\gamma} (t)\, dt
	= 
\frac{s^{\beta}}{(1+s^{-\alpha})^\gamma} =	
	\frac{s^{\alpha\gamma-\beta}}{(s^{\alpha}+1)^{\gamma}}\,, 
	\end{equation}
with $\Re(s)>0, \; |s^\alpha|>1$,
provide the response function and the complex susceptibility ($s = -i\omega$)
respectively, 
  found in the most common  models for non--Debye (or anomalous) relaxation  in dielectrics.
In fact,    the classical dielectric models, referred to as 
Cole--Cole \cite{Cole-Cole1941,Cole-Cole1942},
 Davidson--Cole \cite{Davidson-Cole1951} and 
 Havriliak--Negami \cite{Havriliak-Negami1966,Havriliak-Negami1967},
 are derived according to the scheme 
 \begin{equation} \label{eq:SCHEME}
\alpha \gamma = \beta\; \hbox{with} \; \left\{
\begin{array}{lllll}
{\ds 0<\alpha<1\,, \, \gamma=1} &&& {\mbox{C-C}} \; \{\alpha\}\,,\\
{\ds \alpha=1\,, \, 0<\gamma<1} &&& {\mbox{D-C}} \; \{\gamma\}\,,\\
{\ds 0<\alpha<1\,, \,\, 0< \gamma<1} &&& {\mbox{H-N}} \; \{\alpha,\gamma\}\,.\\
\end{array} \right.
\end{equation}

Moreover, the application of fractional operators based on the Prabhakar function have been recently investigated \cite{GarraGorenfloPolitoTomovski2014} for describing the unsaturated behavior of the free electron laser and the derivation of a renewal point process generalizing the classical homogeneous and the time--fractional Poisson processes.

  In \cite{OliveiraMainardiVaz2011} the authors have proved that the function
 $e_{\alpha,\beta}^{\gamma} (t)$ is locally integrable and completely monotone (LICM)\footnote{%%
 Let us recall that a real function $u(t)$ defined  for $t \in {\mathbb{R}}^+$ 
is said to be completely monotonic (CM)
if it possesses derivatives $u^{(n)}(t)$ for all $n = 0,1,2,3,. .$  and if
$(-1)^n  u^{(n)}(t) \ge 0$
for all 
$t > 0$.
The limit $u^{(n)} (0^+)  = {\displaystyle \lim_{t\to 0^+}} u^{( n)}(t)$ 
finite or infinite exists.
It is known from the Bernstein theorem  that a necessary and sufficient condition that  $u(t)$  be CM is that 
$$ u(t) = \int_0^\infty \hbox{e}^{-rt} \, \hbox{d}\mu(r)\,,$$ 
where $\mu(t)$ is non-decreasing and the integral converges for $ 0 < t < \infty$. 
 In other words  $u(t)$ is required to be the real LT of a non negative measure, in particular 
$$ u(t) =  \int_0^\infty \hbox{e}^{-rt} \, K(r)\, \hbox{d} r\,, \quad K(r) \ge 0\,,$$
where $K(r)$ is a standard or generalized function known as spectral distribution.
For more mathematical details, consult {\it e.g.} the survey by Miller and Samko 
\cite{Miller-Samko2001}.}
  under the conditions
  \begin{equation}
  0<\alpha \le 1, \quad 0<\alpha \gamma \le \beta \le 1.
  \end{equation}
  
As discussed by several authors (e.g., see \cite{Hanyga2005b}) the CM is an essential property for the physical acceptability and realizability of the models since it ensures, for instance, that in isolated systems the energy decays monotonically as expected from physical considerations. Studying the conditions under which the response function of a system is CM is therefore of fundamental importance.

  The purpose of this note is to complement the analysis in 
  \cite{OliveiraMainardiVaz2011} by providing  a more direct proof of the complete monotonicity through an explicit formula for the corresponding spectral density (see Section \ref{S:SpectralDistribution}). Moreover, we discuss the problem of the numerical evaluation of the Prabhakar function and we present some new numerical tests, based on a novel Matlab routine, for validating the results on the CM and better presenting the behavior of the Prabhakar function (see Section \ref{S:NumericalExperiments}). The present analysis allows us to point out that the Havriliak--Negami model described in  (\ref{eq:SCHEME})  may be extended to  $\gamma >1$ provided $\gamma< 1/\alpha$ so that the corresponding response function keeps its LICM character with a non-negative  spectrum density. Finally, section \ref{S:ConcludingRemarks}  is devoted to conclusions and final remarks.

\section{The spectral distribution of the Prabhakar function: analytical aspects}\label{S:SpectralDistribution}

\noindent
We first note that for $0<\alpha<1$ the LT in Eq. (1.3) exhibits  a branch cut on the negative real semi-axis but no poles. Therefore, the inversion of the LT through the Bromwich integral reduces to the evaluation of the integral on an equivalent Hankel path which starts from $-\infty$ along the lower negative real axis, encircles the small circle $|s|=\varepsilon$ in the positive (counterclockwise) sense and returns to $-\infty$ along the upper negative real axis. 

As shown in  \cite{OliveiraMainardiVaz2011}, when $\varepsilon\to0$ it is 
\begin{equation}\label{eq:ML3_LTSpectralDensity}
  e_{\alpha,\beta}^{\gamma} (t) = \int_{0}^{\infty} \e^{-rt} \,
  K_{\alpha,\beta}^{\gamma}(r) \, \dd r\,,
\end{equation}
where
\begin{equation}\label{eq:ML3_TitchmarshFormula}  
   K_{\alpha,\beta}^{\gamma}(r) = 
  \mp \frac{1}{\pi}\,
\hbox{Im} \left[\left. 
{\mathcal E}_{\alpha,\beta}^\gamma (s) \right|_{s=r e^{\pm i\pi}}\right]
\end{equation}
denotes the spectral distribution of $e_{\alpha,\beta}^{\gamma} (t)$. In other words, since ${\mathcal E}_{\alpha,\beta}^{\gamma}(s)$ is required to be the iterated LT of $K_{\alpha,\beta}^\gamma(r)$, we recognize that it is the Stieltjes transform of the spectral distribution.
As a consequence, the spectral distribution can be determined as the inverse Stieltjes transform of 
 ${\mathcal E}_{\alpha,\beta}(s)$ via the so-called Titchmarsh inversion formula (see {\it e.g.} \cite {Titchmarsh1937,Widder1946}) as pointed out in the above equation.

By virtue of the Bernstein theorem, to ensure the complete monotonicity of $e_{\alpha,\beta}^{\gamma} (t)$ the spectral distribution $K_{\alpha,\beta}^\gamma(r)$ is required to be non-negative for all $r\ge 0$.

In \cite{OliveiraMainardiVaz2011} the authors have provided the conditions of complete monotonicity based on the requirements stated in the treatise by Gripenberg et al. \cite{Gripenberg-et-al1990}, see  Theorem 2.6, pp. 144-145,
which provide necessary and sufficient conditions to ensure the CM of a locally integrable function from its LT.
Here we do not take advantage of that theorem but we prefer to compute explicitly
the spectral distribution from the Titchmarsh formula and derive the conditions of non-negativity.

It is indeed possible to observe that 
\begin{equation}
\begin{array}{ll}
K_{\alpha,\beta}^{\gamma}(r) 
	&= 
	{\ds \frac{r^{-\beta}}{\pi} \,{\mbox{Im}} \left\{ 
	{\mbox{e}}^{i\beta\pi} \left( \frac{r^{\alpha} + {\mbox{e}}^{-i \alpha\pi}}
	{r^{\alpha} + 2 \cos ( \alpha\pi)  + r^{-\alpha}}\right)^{\gamma} \right\}} \\ 
	&= 
	{ \ds - \frac{r^{\alpha \gamma -\beta}}{\pi}\, {\mbox{Im}} \left\{
	\frac{{\mbox{e}}^{i (\alpha\gamma-\beta)\pi}}
	{(r^\alpha {\mbox{e}}^{i\alpha\pi}+1)^{\gamma}}
	\right\}}  
\end{array}
\end{equation}
from which, after standard manipulations in complex analysis, we get
\begin{equation}\label{eq:ML3_SpectralDensity}
	K_{\alpha,\beta}^{\gamma} (r) 
	= 
{\ds 
\frac{r^{\alpha\gamma - \beta}}{\pi}\,
 \frac{\sin\left[ \gamma \, \theta_\alpha(r)  + (\beta - \alpha\gamma)\pi \right]}
 {\left( r^{2\alpha} + 2 r^{\alpha}\,\cos(\alpha\pi)+1 \right)^{\gamma/2}}
}\,,  
\end{equation}
where
\begin{equation}
\theta_\alpha(r) : = 
\arctan\left[ \frac{r^\alpha \sin (\pi \alpha)}{r^\alpha \cos(\pi \alpha) +1} \right] 
\in [0,\pi]\,.
\end{equation}
For details we refer to the Appendix where a warning on the correct branch  of the function $\arctan$ is also outlined.

We easily recognize that $\theta_\alpha(r)$ is a non-negative and increasing function of 
$r$ limited by $\alpha \pi \le \pi$,  as shown in the Figure 1, where the dotted lines indicate the limit values  $\alpha\pi$.
In fact for  $r \gg 1 $ it is direct to check its asymptotic behavior
\begin{equation}
\frac{r^\alpha \sin (\pi \alpha)}{r^\alpha \cos(\pi \alpha) +1} = \frac{ \sin (\pi \alpha)}{\cos(\pi \alpha) + 1/r^\alpha}
\le \frac{ \sin (\pi \alpha)}{\cos(\pi \alpha)} = \tan (\pi \alpha) \,.
\end{equation}
\begin{figure}%[p!] 
\centering
\includegraphics[width=0.7\textwidth]{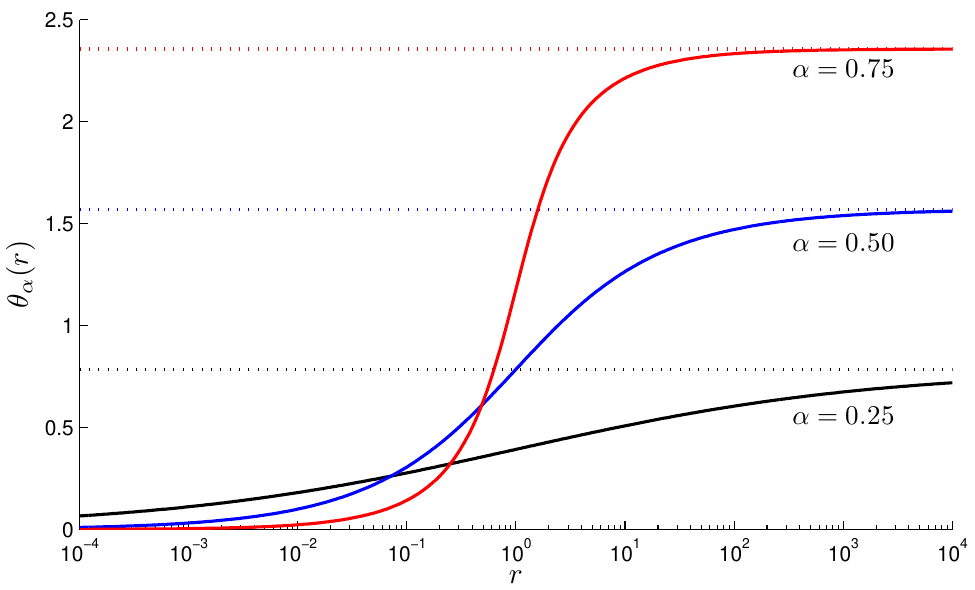}
\caption{The function $\ds \theta_\alpha (r)$ for $\ds \alpha = 0.25$, $\ds \alpha = 0.50$ and $\ds \alpha = 0.75$.}\label{theta}
\end{figure}

Then we recognize that in order the  spectral distribution to be non-negative the argument of the $\sin$ function in the numerator must be included in the closed interval $[0, \pi]$ and henceforth
we find the conditions stated in \cite{OliveiraMainardiVaz2011}, that we repeat for convenience:
 
\begin{equation}\label{eq:ConditionCM}
  0<\alpha \le 1, \quad 0<\alpha \gamma \le \beta \le 1.
 \end{equation}

\begin{figure}[ht!]
\centering
\includegraphics[width=0.7\textwidth]{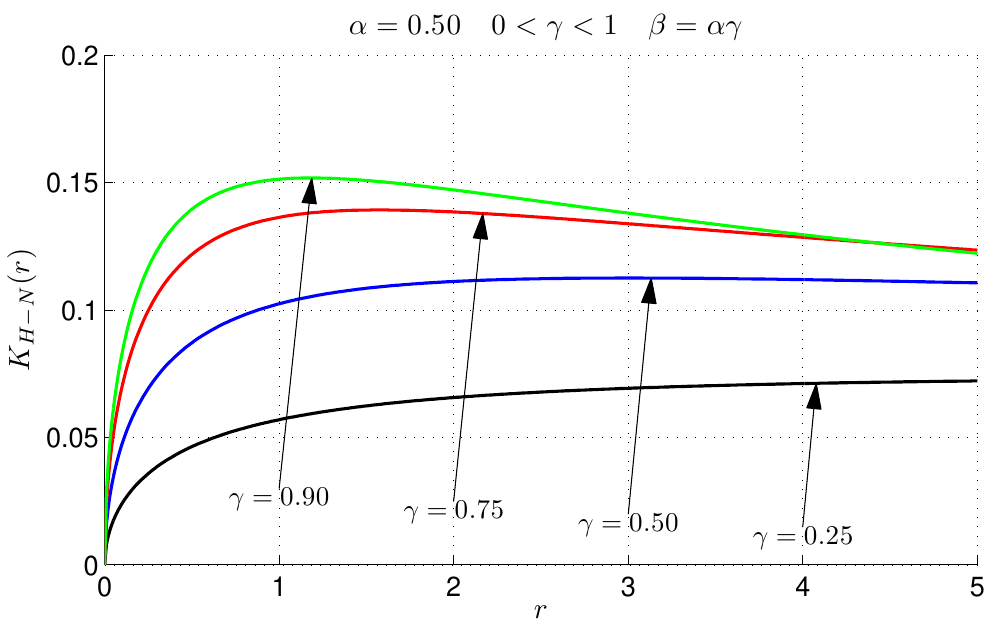}
\caption{The spectral distribution for the Havriliak-Negami model 
 $ \ds K_{{\mbox{H-N}}}(r) $ 
for  $ \alpha = 0.5$ and $0<\gamma<1$.} 
\label{fig:Fig_HKP_alpha050_lower1}
\end{figure}

\begin{figure}[ht!] 
\centering
\includegraphics[width=0.7\textwidth]{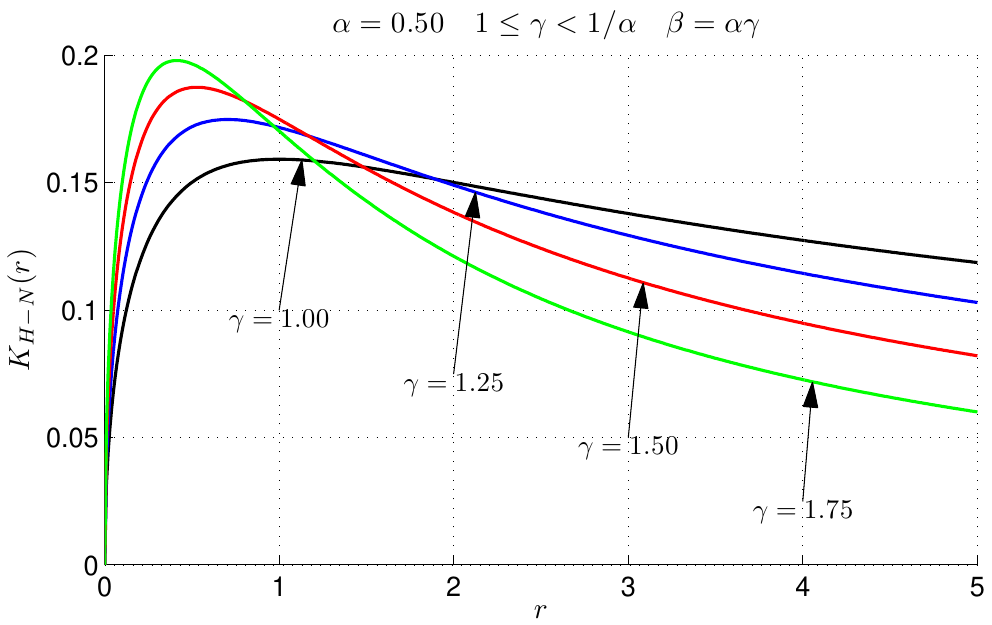}
\caption{The spectral distribution for the Havriliak-Negami model 
$ \ds K_{{\mbox{H-N}}}(r)$, for  
$\alpha=0.5$, and $ 1\le \gamma < 2=1/\alpha$.}
\label{fig:Fig_HKP_alpha050_great1}
\end{figure}

\begin{figure}[ht!]
\centering
\includegraphics[width=0.7\textwidth]{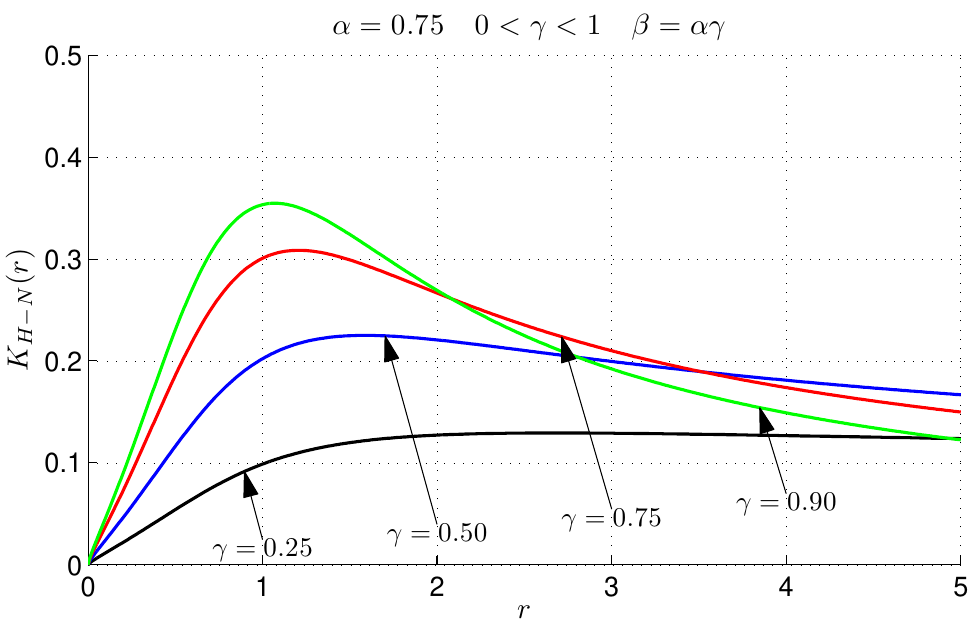}
\caption{The spectral distribution for the Havriliak-Negami model,
 $ \ds K_{{\mbox{H-N}}}(r) $, for $\alpha =0.75$ and $ 0<\gamma<1$.}\label{fig:Fig_HKP_alpha075_lower1}
\end{figure}

\begin{figure}[ht!] 
\centering
\includegraphics[width=0.7\textwidth]{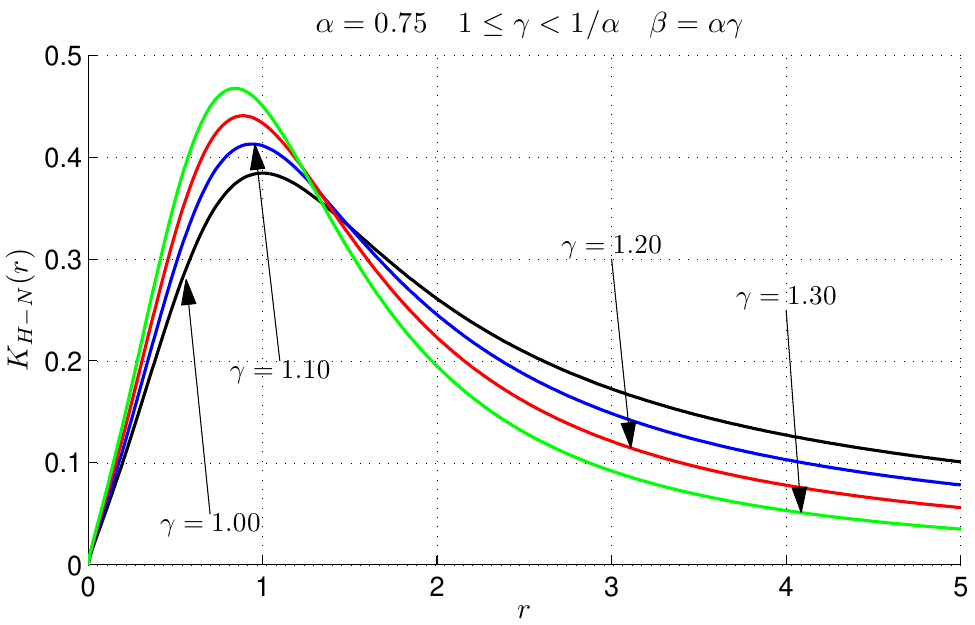}
\caption{The spectral distribution for the Havriliak-Negami model 
$ \ds K_{{\mbox{H-N}}}(r)$, for $\alpha=0.75$ and
 $1\le \gamma < 4/3 = 1/\alpha $.}\label{fig:Fig_HKP_alpha075_great1}
\end{figure}

In Figures \ref{fig:Fig_HKP_alpha050_lower1}, \ref{fig:Fig_HKP_alpha050_great1}, \ref{fig:Fig_HKP_alpha075_lower1} and \ref{fig:Fig_HKP_alpha075_great1} we report  the spectral distributions for the H-N model,
obtained from Eq. (\ref{eq:ConditionCM}) 
assuming $\beta =\alpha \gamma$, keeping fixed $\alpha\in\{0.5,0.75\}$ and varying $\gamma$.
 We explicitly note the non-negativity of the spectral distributions  
(ensuring the complete monotonicity of the corresponding response function) 
not only when  $\gamma \le 1$ but more generally when  $\alpha\gamma<1$, so 
$\gamma$ can overcome the value 1 and generalize the H-N model. 

%%%%%%%%%%%%%
%%%%%%%%%%%%
%\newpage
 %%%%%%%%

\section{Numerical computation and validation of theoretical results}\label{S:NumericalExperiments}

The aim of this section is to validate the theoretical results on the CM of the Prabhakar function by means of some numerical experiments. 

We will first discuss some methods for evaluating the function and we will validate their results by comparison with some asymptotic expansions; hence we will provide some graphical evidences in order to confirm the correctness of the condition stated in (\ref{eq:ConditionCM}) for the CM of the function under investigation.

\subsection{Numerical computation of the Prabhakar function}

To the best of our knowledge, very few mathematical packages provide built-in facilities for the computation of the ML function. One of the most used codes, the Matlab {\tt mlf.m} function devised by Podlubny and Kacenak \cite{PodlubnyKacenak2012} and implementing algorithms similar to those investigated in \cite{GorenfloLoutchkoLuchko2002}, allows to compute the two parameters function with high accuracy but does not apply to the more general three parameters case. This is also the case of the computational software Mathematica in which, starting from the recent version 9, the ML function has been introduced but only for 1 and 2 parameters. Other algorithms for the computation of the 2 parameters ML function were developed in \cite{SeyboldHilfer2005,HilferSeybold2006,SeyboldHilfer2008}.

For the computation of the Prabhakar function it is therefore necessary to device specific algorithms and we consider here three different approaches:
\begin{enumerate}
	\item truncation of the series expansion in (\ref{eq:ML3_SeriesExpansion});
	\item evaluation of the LT (\ref{eq:ML3_LTSpectralDensity}) of the spectral distribution (\ref{eq:ML3_SpectralDensity});
	\item numerical inversion of the LT (\ref{eq:ML3_LaplaceTransform}).
\end{enumerate}

In the first approach some terms of the series expansion in the original definition (\ref{eq:ML3_SeriesExpansion}) are evaluated and summed up until their difference in absolute value is below a given threshold. This is the most straightforward and easy to implement approach but, unfortunately, it is the less reliable. It indeed works in an acceptable way only for small values of $t$ since otherwise the slow convergence imposes the computation of a huge number of terms; in this case, not only the computation requires a large amount of time but it can also be impossible due to overflow errors in computing $k!$ and $\Gamma(\alpha k + \beta)$ and numerical cancellation in summing terms with very different magnitude.

With the second approach, the ML function (\ref{eq:ML3_SeriesExpansion}) is evaluated as the LT of its spectral distribution (\ref{eq:ML3_SpectralDensity}) by applying some numerical quadrature rule (for instance, we use the adaptive Gauss--Kronrod quadrature implemented by the Matlab {\tt quadgk} function supporting infinite intervals and functions with endpoints singularities). We must however observe that this kind of quadrature is usually very time--consuming and moreover, as we will experimentally observe, it works only with singularities of moderate type, i.e. when $\beta$ is not much larger than $\alpha \gamma$.

The third approach is performed into two main steps: the Bromwich line in the formula for the inversion of the LT is first deformed into an equivalent path beginning and ending in the left half of the complex plane (in this way the exponential factor 
$\e^{st}$ rapidly decays without strong oscillations which are source of numerical instability); hence a numerical quadrature (for instance, the simple compound trapezoidal rule) is applied along this path. An algorithm based on contours of parabolic type has been extensively studied in \cite{WeidemanTrefethen2007,TrefethenWeideman2014} and successfully applied to the two parameters ML function in \cite{GarrappaPopolizio2013}; it is appreciated since any preassigned accuracy can be achieved with a modest computational effort; moreover, since it is mainly based on the analysis of the singularities of the LT of the ML function, it can be easily adapted to evaluate the three parameters extension too \cite{Garrappa2015}.

While the first two approaches apply to a very restricted range of parameters and arguments, the last one appears as the most robust and reliable for general use. We thus intend to evaluate the ML function by inverting its LT and use the other two methods just for validating the obtained results.

\begin{figure}[ht]
	\centering
	\includegraphics[width=0.7\textwidth]{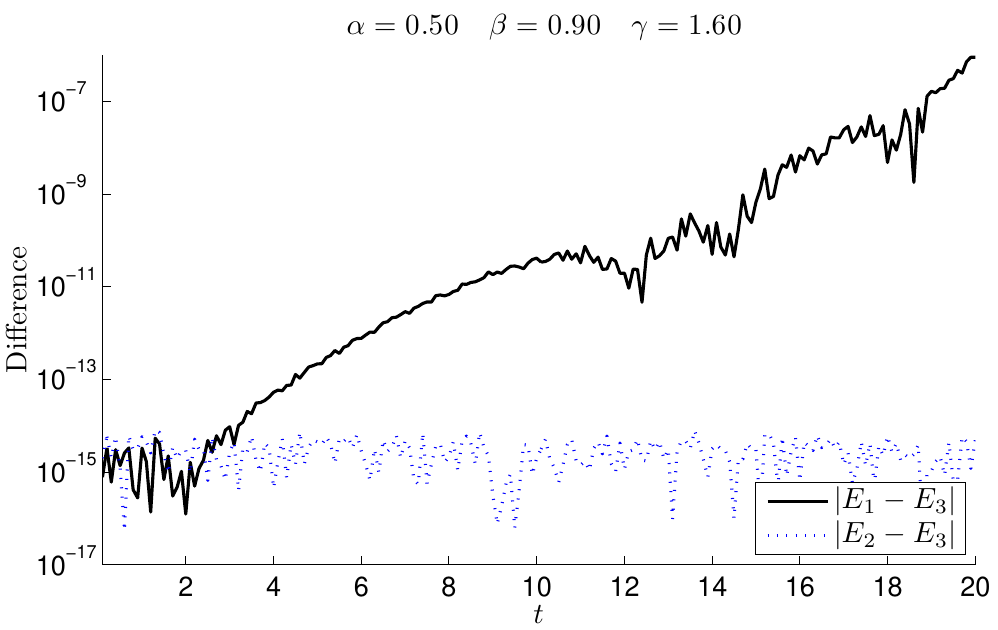} 
	\caption{Comparison of different approaches for evaluating the three parameters ML function.}\label{fig:Fig_alpha050_beta090_gamma160_Accuracy}
\end{figure}

\begin{figure}[ht]
	\centering
	\includegraphics[width=0.7\textwidth]{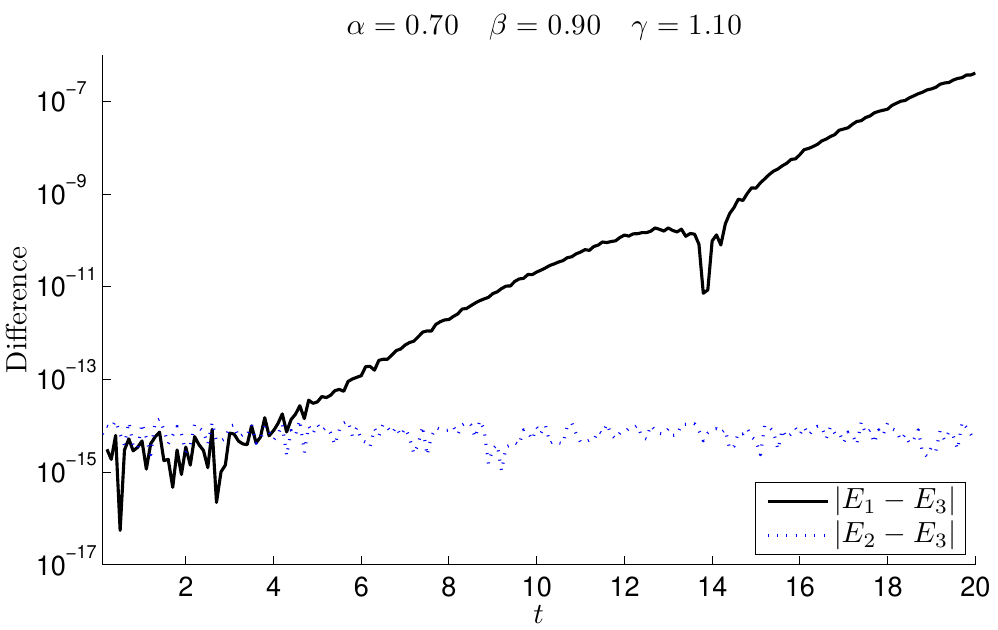} 
	\caption{Comparison of different approaches for evaluating the three parameters ML function.}\label{fig:Fig_alpha070_beta090_gamma110_Accuracy}
\end{figure}

As we can see from Figures \ref{fig:Fig_alpha050_beta090_gamma160_Accuracy} and \ref{fig:Fig_alpha070_beta090_gamma110_Accuracy}, where we have plotted the difference, in absolute values, between the outcomes resulting from the application of the three above described approaches (labeled respectively as $E_1$,$E_2$ and $E_3$), the method based on the truncation of the series expansion of the ML function is reliable only when $t$ is not too far from the origin, otherwise large errors are expected. The evaluation by the LT of the spectral distribution and the numerical inversion of the LT of the ML function provide very similar results, with a difference very close to the machine precision, thus confirming the high accuracy which they are able to provide.

We must however observe that the numerical evaluation of the LT (\ref{eq:ML3_LTSpectralDensity}) is a quite complicated task when $\beta > \alpha \gamma$ because of the endpoint singularity in the spectral distribution (\ref{eq:ML3_SpectralDensity}). The Matlab quadrature code  {\tt quadgk} used for approximating the integral in (\ref{eq:ML3_LTSpectralDensity}) has been able to provide accurate results when $\alpha=0.5$, $\beta=0.9$ and $\gamma=1.6$ (see Figure \ref{fig:Fig_alpha050_beta090_gamma160_Accuracy}), i.e. for a singularity with a strength $\alpha\gamma - \beta = 0.1$ but not when $\beta=0.97$, i.e. when the strength of the singularity is $0.17$. At the same time it is has been possible to correctly evaluate the integral for $\alpha=0.7$, $\beta=0.9$ and $\gamma=1.1$ as shown in Figure \ref{fig:Fig_alpha070_beta090_gamma110_Accuracy} but not when $\beta=0.95$.

We think however that this approach deserves a more in-depth investigation (which is beyond the scope of this paper) since more suitable quadrature rules could be applied to efficiently evaluate (\ref{eq:ML3_LTSpectralDensity}).

\subsection{Asymptotic expansion}

To further validate the numerical computation of the Prabhakar function, especially for large values of the argument $t$, it is useful to derive some asymptotic expansions for $t\to\infty$.

The first expansions in the whole complex plane were provided by Wright in 1935 and 1940 in two distinct papers \cite{Wright1935,Wright1940} devoted to the analysis of more general hypergeometric functions. More recently, Paris \cite{Paris2010} improved the analysis of Wright by considering also exponentially small expansions.

Although the results in the papers or Wright and Paris are very rigorous, their practical use is quite complicate since it demands for the computation of some residues of non elementary functions; thus, for our aims it is more convenient to derive specific expansions by inverting the LT (\ref{eq:ML3_LaplaceTransform}) as $s\to 0$. 

It is simple to observe that the full series expansion in positive powers of $s$ of 
$(s^\alpha +1)^{-\gamma}$ is
\begin{equation}\label{eq:ML3_Asymptotic_LT}
	(s^\alpha +1)^{-\gamma} =
		1 + \sum_{k=1}^\infty \binom{-\gamma}{k} \, s^{\alpha k}
\end{equation}
where the binomial coefficients can be computed by recurrence and are related to Gamma functions as follows
\begin{equation}
 \binom{-\gamma}{k} =
 \frac{\Gamma(-\gamma +1)}{\Gamma(-\gamma -k+1)\,k!} =
 (-1)^k \, \frac{\Gamma(\gamma +k)}{\Gamma(\gamma)\,k!} .
\end{equation}

By term by term inversion, and recalling the generalized LT pair (see, for instance, the classical book of Doetsch \cite{Doetsch1974})
\begin{equation}
	s^\nu \div  \frac{t^{-\nu -1}}{\Gamma (-\nu)} 
\end{equation}
which holds for non integer $\nu >0$, we get as $t \to \infty$ the following asymptotic series for $\alpha \gamma -\beta \ne 0$
\begin{equation}\label{eq:ML3_AsymptoticExpansion1}
	{\mathcal E}_{\alpha,\beta}^\gamma (s) = 
		\frac{s^{\alpha\gamma-\beta}}{(s^\alpha+1)^{\gamma}}
	\div 
	e_{\alpha,\beta}^\gamma (t) = 
		t^{\beta- \alpha \gamma -1}
	  \sum_{k=0}^\infty \binom{-\gamma}{k}  
	    \frac{t^{-\alpha k}}{\Gamma(\beta -\alpha\gamma-\alpha k)}  ,
\end{equation}
whilst, by neglecting in (\ref{eq:ML3_Asymptotic_LT}) the spurious singular term $1 \div \delta(t)$,  in the case $\alpha \gamma -\beta =0$ we have 
\begin{equation}\label{eq:ML3_AsymptoticExpansion2}
	{\mathcal E}_{\alpha,\beta}^\gamma (s) = 
	\frac{1}{(s^{\alpha}+1)^{\gamma}}
	\div  
		e_{\alpha,\beta}^\gamma (t) = 
			\sum_{k=1}^\infty \binom{-\gamma}{k}  \frac{t^{-\alpha k -1}}{\Gamma(-\alpha k)}
\end{equation}
(similar results are obtained, by using the Mellin--Barnes integral representation of the Prabhakar function, in \cite{HauboldMathaiSaxena2004} where however, perhaps due to a misprint, it seems that a term $(-1)^k$ is missing). As a consequence, the dominant term for $t\to\infty$ (asymptotic representation) is given by

\begin{equation}
 e_{\alpha,\beta}^{\gamma} (t) \sim
	\left\{ \begin{array}{ll}
		{\ds \frac{t^{\beta- \alpha \gamma -1}}{\Gamma(\beta -\alpha\gamma)}} \,,
		& \hbox{if} \; 0< \alpha \gamma <\beta \le 1\,, \\
		{\ds - \gamma \frac{t^{- \alpha  -1}}{\Gamma(- \alpha)}} \,,
		& \hbox{if} \;  0< \alpha \gamma =\beta \le 1\,.
	\end{array}\right. 
\end{equation}

\begin{figure}[ht]
	\centering
	\includegraphics[width=0.8\textwidth]{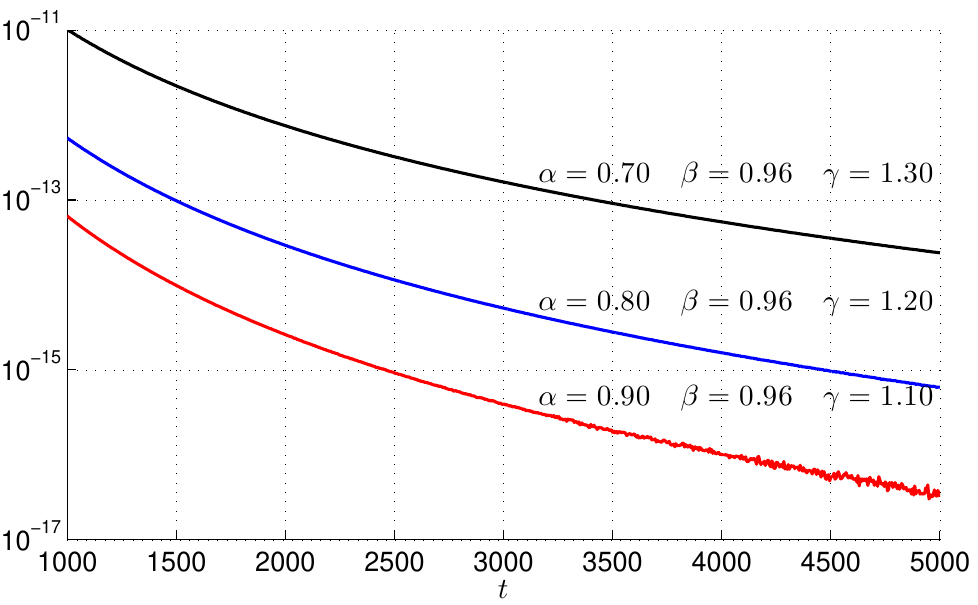} 
	\caption{Difference between the 3--parameters ML function evaluated by numerical inversion of the LT and the asymptotic expansions (\ref{eq:ML3_AsymptoticExpansion1}-\ref{eq:ML3_AsymptoticExpansion2}).}\label{fig:Fig_Difference_Asymptotic}
\end{figure}

The plot in Figure \ref{fig:Fig_Difference_Asymptotic} presents the difference, in absolute value, between the ML function (\ref{eq:ML3_Generalized}) evaluated by numerically inverting the LT and the truncation (after the first 3 terms) of the asymptotic expansions (\ref{eq:ML3_AsymptoticExpansion1}-\ref{eq:ML3_AsymptoticExpansion2}); since the expansion are derived for $t\to\infty$, the first part of the positive real axis is omitted in the plot.

As we can see, for the three selections of the order parameters $\{\alpha, \beta, \gamma\}$ (which cover the three main different cases $\alpha\gamma < \beta$, $\alpha \gamma = \beta$ and $\alpha \gamma > \beta$) the numerical evaluation of the ML function (\ref{eq:ML3_Generalized}) agrees in a reasonable way with the expansions (\ref{eq:ML3_AsymptoticExpansion1}-\ref{eq:ML3_AsymptoticExpansion2}) and their difference is very small (in some cases it rapidly achieves values below the machine precision of the floating--point arithmetic used for the computation). 

The rate at which the asymptotic expansions (\ref{eq:ML3_AsymptoticExpansion1}-\ref{eq:ML3_AsymptoticExpansion2}) converge to the ML function (\ref{eq:ML3_Generalized}) obviously depends on $\beta- \alpha \gamma -1$ (or $-\alpha-1$, when $\alpha\gamma=\beta$), i.e. the power of the leading term in the expansion of $e_{\alpha,\beta}^{\gamma}(t)$ as $t\to\infty$.

\subsection{Verifying the conditions for CM}

On the basis of the observations in the previous subsections, the method exploiting the numerical inversion of the LT tuns out to be reliable and efficient for the computation of the Prabhakar function and therefore it will be used in this work for validating the theoretical findings on CM properties.

\begin{figure}[ht]
	\centering
	\includegraphics[width=0.8\textwidth]{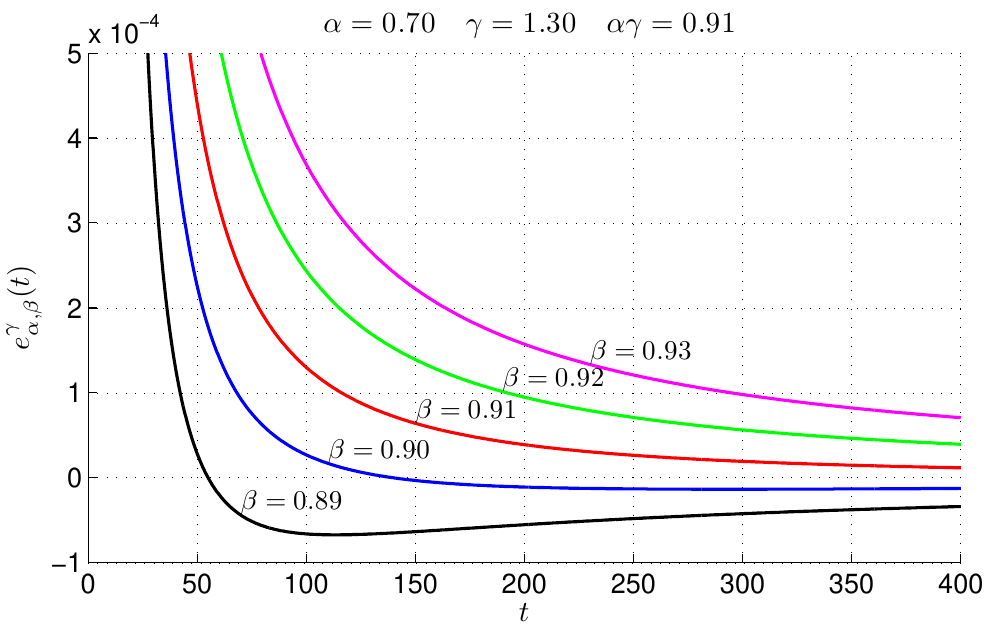} 
	\caption{Graph of the ML function for $\alpha =0.70$, $\gamma=1.30$ and $\beta$ varying.}\label{fig:Fig_alpha0.70_gamma1.30_E_beta}
\end{figure}

In Figure \ref{fig:Fig_alpha0.70_gamma1.30_E_beta} we show, for $\alpha =0.70$ and $\gamma=1.30$ the behavior of the function as the second parameter, namely $\beta$, varies in a neighborhood of the threshold $\alpha\gamma = 0.91$ for which CM is expected.

As we can clearly see, when $\beta < 0.91$ negative values are obtained; the non monotonic character, which can be hardly observed for $\beta=0.90$, is more manifest for $\beta = 0.89$.

An additional test can be performed on the derivatives of $e_{\alpha,\beta}^{\gamma}(t)$; after direct differentiation of each term in the series defining the ML function (\ref{eq:ML3_Generalized}), it is indeed easy to verify that
\begin{equation}\label{eq:DerivativeML3}
	\frac{d^k}{d t^{k}} e_{\alpha,\beta}^{\gamma}(t) = e_{\alpha,\beta-k}^{\gamma}(t) 
	, \quad t > 0.
\end{equation}

\begin{figure}[ht]
	\centering
	\includegraphics[width=0.8\textwidth]{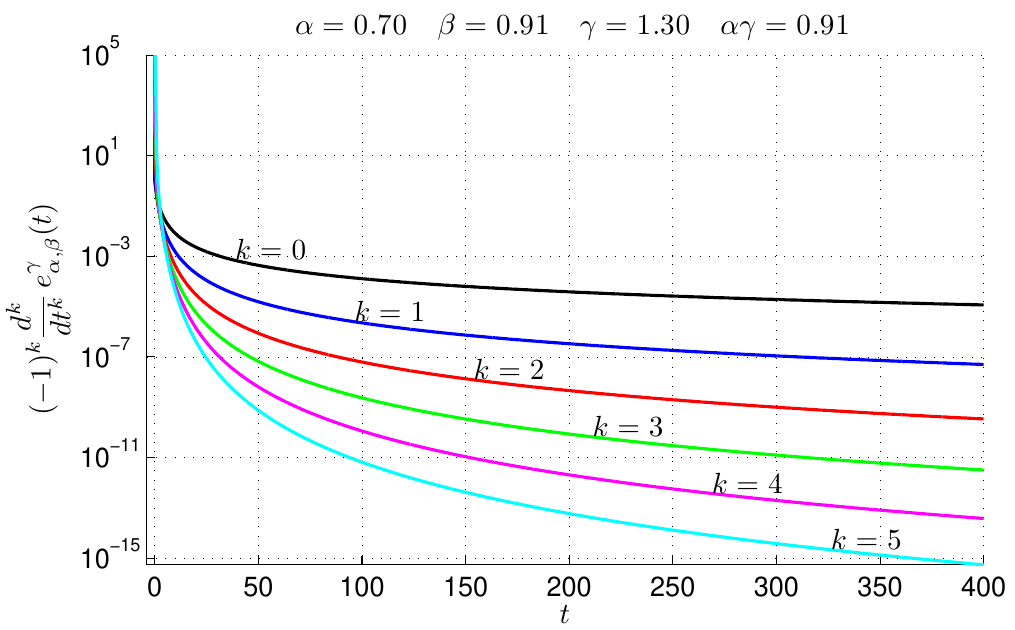} 
	\caption{First derivatives of the ML function for $\alpha =0.70$, $\beta=0.91$ and $\gamma=1.30$.}\label{fig:Fig_alpha070_beta091_gamma130_E_log}
\end{figure}

From Figure \ref{fig:Fig_alpha070_beta091_gamma130_E_log}, where the first few derivatives (with alternating signs) $(-1)^{k} \frac{d^k}{d t^{k}} e_{\alpha,\beta}^{\gamma}(t)$, $k=0,\dots,5$, are plotted by means of (\ref{eq:DerivativeML3}), we can clearly observe that, as expected from theoretical predictions, non negative values are obtained.

%As we can clearly see in Figure \ref{fig:Fig_alpha070_beta091_gamma130_E_log}, where the derivatives (with alternating signs) $(-1)^{k} \frac{d^k}{d t^{k}} e_{\alpha,\beta}^{\gamma}(t)$, $k=0,\dots,5$, are plotted by means of (\ref{eq:DerivativeML3}), non negative values converging to 0 as $t \to \infty$ are obtained as expected from  theoretical predictions.

%As we can see from the plots of $(-1)^{k} \frac{d^k}{d t^{k}} e_{\alpha,\beta}^{\gamma}(t)$, $k=0,\dots,5$, in Figure \ref{fig:Fig_alpha070_beta091_gamma130_E_log} all the first derivatives tend to 0 as $t\to \infty$ but they are all non negative as expected from the theoretical analysis.

\begin{figure}[ht]
	\centering
	\includegraphics[width=0.8\textwidth]{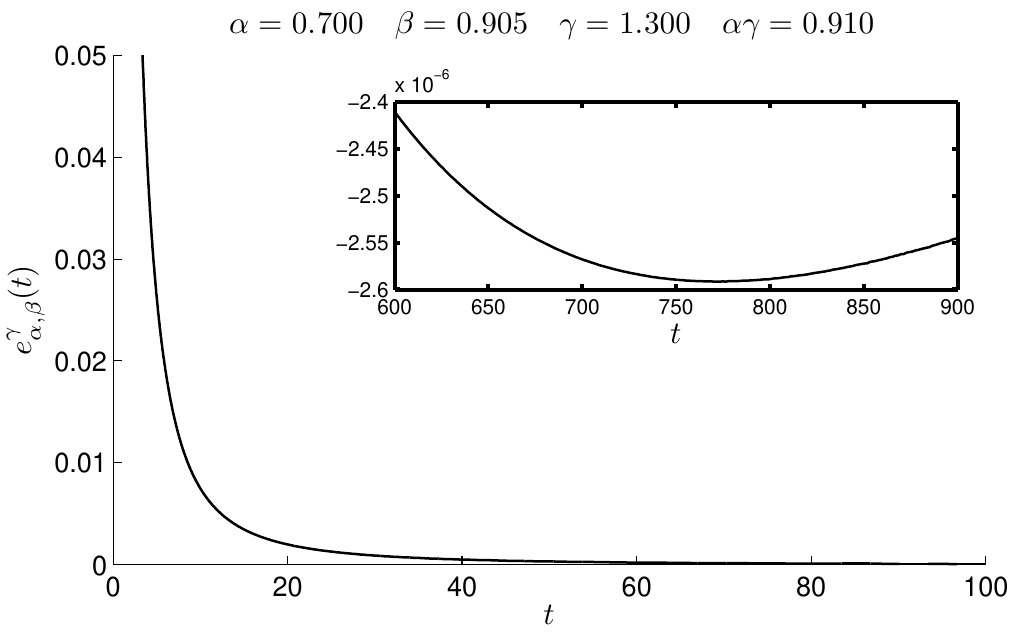} 
	\caption{Apparent non-negativity and monotonicity for $\alpha =0.700$, $\beta=0.905$ and $\gamma=1.300$.}\label{fig:Fig_alpha070_beta0905_gamma130_E}
\end{figure}

It is interesting to note that for values of $\beta$ below but very close to the threshold values $\alpha \gamma$ for which CM is no longer assured, the CM can be only apparent. As we can see in Figure \ref{fig:Fig_alpha070_beta0905_gamma130_E} indeed it could seem that the function is non negative and monotonic but a closer inspection, and on a wider range for $t$, shows that negative values and a non monotonic behavior are achieved (see the box inside the same Figure \ref{fig:Fig_alpha070_beta0905_gamma130_E}).

\begin{figure}[ht]
	\centering
	\includegraphics[width=0.8\textwidth]{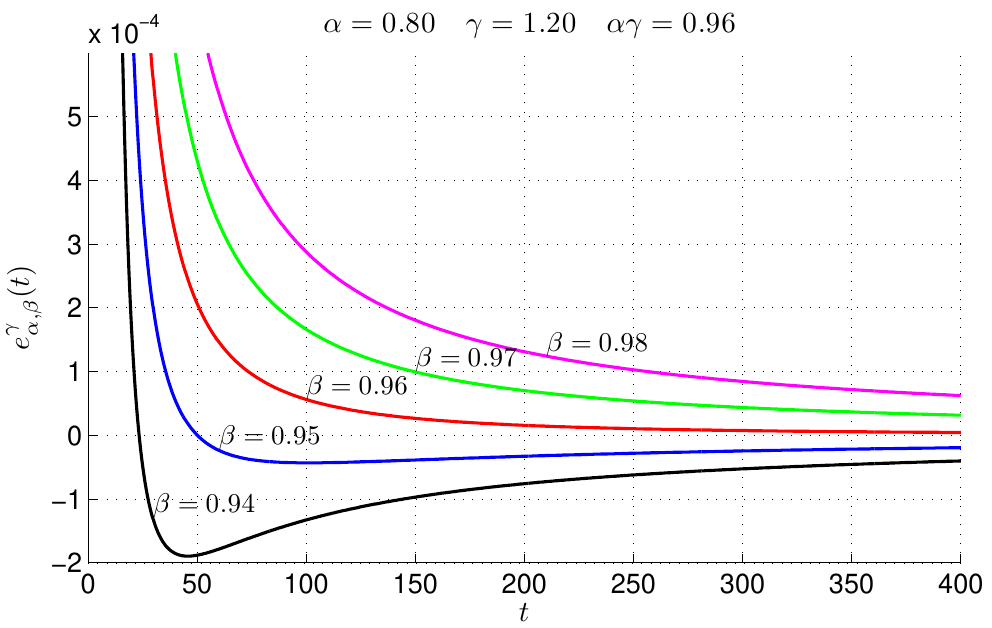} 
	\caption{Graphs (as $\beta$ varies) of the ML function for $\alpha =0.80$ and $\gamma=1.20$.}\label{fig:Fig_alpha0.80_gamma1.20_E_beta}
\end{figure}

We conclude this Section by presenting the plots of the Prebhakar function (\ref{eq:ML3_Generalized}) for other instances of the parameters $\alpha$ and $\beta$. In particular in Figure \ref{fig:Fig_alpha0.80_gamma1.20_E_beta} it is presented the case $\alpha=0.80$ and $\gamma=1.20$ for which CM is expected for $\beta > 0.96$.

\begin{figure}[ht]
	\centering
	\includegraphics[width=0.8\textwidth]{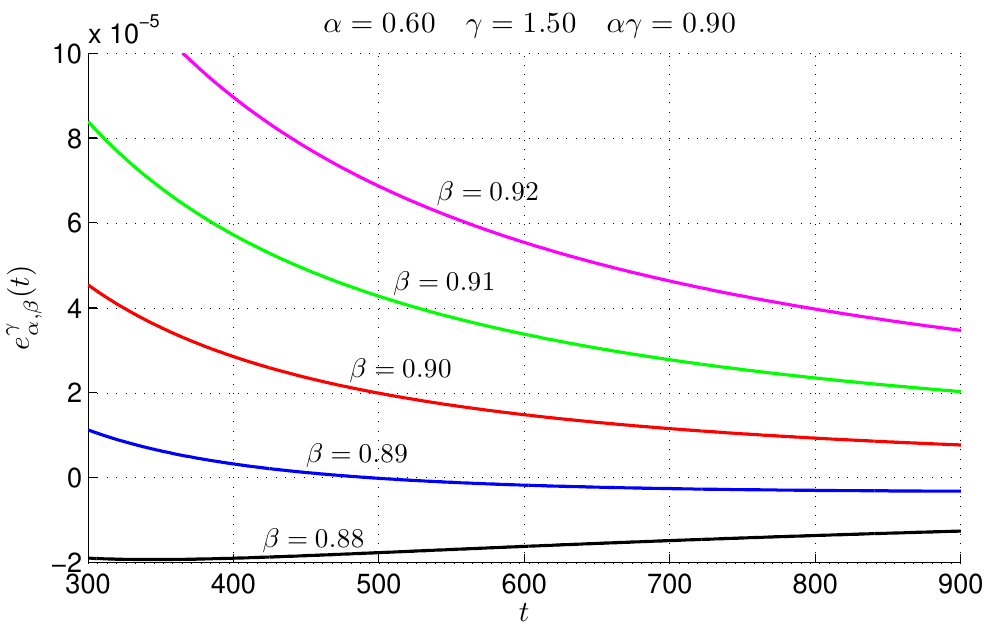} 
	\caption{Graphs (as $\beta$ varies) of the ML function for $\alpha =0.60$ and $\gamma=1.5$.}\label{fig:Fig_alpha060_gamma150_E_beta}
\end{figure}

In Figure \ref{fig:Fig_alpha060_gamma150_E_beta} we present the case $\alpha=0.6$ and $\gamma=1.5$ for which the CM is instead expected when $\beta > 0.9$; we show only the interval for $t$ in which the most interesting phenomena are present.

All the plots presented in this Section seem to validate, in a quite clear way, the theoretical findings.

\section{Concluding remarks}\label{S:ConcludingRemarks}

Some properties of the three parameters Mittag-Leffler function have been investigated. In particular, we have established the conditions on the parameters $\alpha$, $\beta$ and $\gamma$ for which the function turns out locally integrable and completely monotonic. 

These conditions are essential in order to suitably model relaxation phenomena of non-Debye type, such as anomalous polarization processes in dielectrics. We think that these results can be of some help especially for parameter identification, a task which is usually performed in an experimental way. In particular we  have been able to  extend the validity of the classical Havriliak-Negami model. 

Moreover, some numerical methods have been discussed and compared with the aim of identifying suitable techniques for the accurate and efficient numerical computation of the three parameters Mittag-Leffler function. An approach based on the inversion of the LT, which appears as the most reliable, has been used for validating the theoretical findings and providing some graphical representations of the function under investigation.

\section*{Acknowledgements}\label{S:Acknowledgements}
The authors appreciate constructive remarks and suggestions of the
referees that helped to improve the manuscript.

The work of F.Mainardi has been carried out in the framework  of the GNFM - INdAM  activities. 

The work of R.Garrappa has been supported under the GNCS - INdAM Project 2014.

%\appendix{The spectral distribution of the Prabahakar function} \label{proof}
\section*{Appendix: The spectral distribution of the Prabahakar function} \label{proof}

For ease of presentation we have collected in this Appendix some details regarding the derivation of the spectral distribution $K_{\alpha,\beta}^{\gamma} (r)$. 

After applying the Titchmarsh inversion formula, from (\ref{eq:ML3_TitchmarshFormula}) we have

$$
\begin{array}{lll}
K_{\alpha,\beta}^{\gamma} (r) 
&= 
{\ds \frac{r^{-\beta}}{\pi} \,{\mbox{Im}} \left\{ 
{\mbox{e}}^{i\beta\pi} \left( \frac{r^{\alpha} + {\mbox{e}}^{-i \alpha\pi}}
{r^{\alpha} + 2 \cos ( \alpha\pi)  + r^{-\alpha}}\right)^{\gamma} \right\}}  \\
&= 
{ \ds - \frac{r^{\alpha \gamma -\beta}}{\pi}\, {\mbox{Im}} \left\{
\frac{{\mbox{e}}^{i (\alpha\gamma-\beta)\pi}}
{(r^\alpha {\mbox{e}}^{i\alpha\pi}+1)^{\gamma}}
\right\}}  \\
&= { \ds - \frac{r^{\alpha \gamma -\beta}}{\pi}\, {\mbox{Im}} \left\{
\frac{{\mbox{e}}^{i (\alpha\gamma-\beta)\pi}}
{(r^\alpha {\mbox{e}}^{i\alpha\pi}+1)^{\gamma}} 	\frac{(r^\alpha {\mbox{e}}^{-i\alpha\pi}+1)^{\gamma}}{(r^\alpha {\mbox{e}}^{-i\alpha\pi}+1)^{\gamma}}
\right\}}
\end{array}
\leqno(A.1)
$$

It is now easy to check that the denominator is real and non-negative, so we set
$$
\xi := (r^\alpha {\mbox{e}}^{i\alpha\pi}+1) (r^\alpha {\mbox{e}}^{-i\alpha\pi}+1) = r^{2\alpha} + 2 r^{\alpha}\,\cos(\alpha\pi)+ 1\,,
\leqno(A.2)$$
with
 $ \xi \ge (r^\alpha -1)^2 \ge 0$
and consequently  $ \xi^{{\gamma}/{2}} \ge 0$. 
For the numerator we set
$$
z :=(r^\alpha {\mbox{e}}^{-i\alpha\pi}+1) =
[r^\alpha  \,\cos(\alpha\pi) + 1] - i\, r^\alpha  \,\sin (\alpha\pi) =  
 \rho\,{\mbox{e}}^{-i\theta}\,,
\leqno(A.3)
$$
where $0\le \theta\le \pi$ (being $0<\alpha\le 1)$.
Then  
$$\rho = |z|= \sqrt{
[{\mbox{Re} (z)}]^2  + [{\mbox{Im} ( z)}]^2 } = \xi^{\frac{1}{2}}
\leqno(A.4)$$ 
and 
$$ \tan \theta = - \frac{ \mbox{Im} (z)}{ \mbox{Re}  (z)}=
\frac{r^\alpha \sin (\pi\alpha)}{r^\alpha \cos(\pi\alpha) +1}\,.
\leqno(A.5)
$$
%%%%%%%55
Using the \textit{de Moivre's formula} 
 $ ( \cos \psi + i\sin \psi ) ^n = \cos (n \psi) + i \sin (n\psi) $ 
 we get:
$$
\begin{array}{lll}
K_{\alpha,\beta}^{\gamma} (r) 
\!\! &= \!\!{ \ds
 - \frac{r^{\alpha \gamma -\beta}}{\pi}  \, 
\frac{ 
 {\mbox{Im}}
  \left\{ \left[\cos(\alpha\gamma-\beta)\pi + i\sin(\alpha\gamma-\beta)\pi \right] \,
   \left[ \cos (\gamma \theta) - i \sin (\gamma \theta)\right]\right\}}
{ \xi^{\gamma/2}}
}
 \\ \\
\!\! &= \!\!{\ds - \frac{r^{\alpha \gamma -\beta}}{\pi} 
\,
\frac{\left[ -\cos(\alpha\gamma-\beta)\pi \,\sin (\gamma \theta)  + \sin(\alpha\gamma-\beta)\pi \,\cos (\gamma \theta) \right]}
{ \xi^{\gamma/2}}
}
 \\ \\
\!\! &= \!\! {\ds
 - \frac{r^{\alpha \gamma -\beta}}{\pi} \, 
\frac{
\sin \left[ (\alpha\gamma - \beta)\pi - \gamma\theta  \right]}
{ \xi^{\gamma/2}}
}
= {\ds 
\frac{r^{\alpha \gamma -\beta}}{\pi}\, 
\frac{\sin \left[  \gamma\theta + (\beta-\alpha\gamma)\pi  \right]}
{ \xi^{\gamma/2}}
}\,,
\end{array}
\leqno(A.6)
$$
where 
$$
\theta =\theta_\alpha(r) : =\arctan \left[ \frac{r^\alpha \sin (\pi\alpha)}{r^\alpha \cos(\pi\alpha) +1} \right]
\in [0, \pi].
\leqno(A.7)
$$
%%as in Eq.(2.20).
% \\
As noted by  Zorn  for the Havriliak-Negami model \cite{Zorn1999}, see 
our analysis for $\alpha \gamma -\beta=0$,
we need to chose the arctangent's value in the interval 
$[0,\pi]$, which is possible if one considers $\arctan (x)$ to be a multivalued function. 
In this sense our proposed formula is always valid if only correctly interpreted. Staying with the usual definition of $\arctan (x)$ as a function into $[-\pi/2,\pi/2]$, one has 
to add $\pi$  to avoid  the negative values instead of the changing of sign.

%\section*{References}
%\bibliographystyle{elsarticle-num}
%\bibliography{ML3_Biblio}

\end{document}